# Hour-scale persistent negative anomaly of atmospheric electrostatic field near the epicenter before earthquake

Tao Chen[1]*, Han Wu[1,2], Chi Wang[1], Xiaoxin Zhang[3], Xiaobin Jin[4], Qiming Ma[5], Jiyao Xu[1], Suping Duan[1], Zhaohai He[1], Hui Li[1], Saiguan Xiao[1], Xizhen Wang[6], Xuhui Shen[7], Quan Guo[7], Ilan Roth[8], Vladimir Makhmutov[9], Yong Liu[1], Jing Luo[1], Xiujie Jiang[1], Lei Dai[1], Xiaodong Peng[1], Xiong Hu[1], Lei Li[1,2], Chen Zeng[1,2], Jiajun Song[5], Fang Xiao[5], Jianguang Guo[3], Cong Wang[3], Hanyin Cui[10], Chao Li[10], Qiang Sun[11].

[1] State Key Laboratory of Space Weather, National Space Science Center, Chinese Academy of Sciences

[2] University of Chinese Academy of Sciences

[3] National Center for Space Weather, Chinese Meorological Administration

[4] Sichuan Meteorology Agency, Chinese Meorological Administration

[5] Institute of Electrical Engineering, Chinese Academy of Sciences

[6] Institute of geophysics, Chinese Earthquake administration

[7] Institute of Crustal Dynamics, Chinese Earthquake administration

[8] Space Science Laboratory, UC Berkeley, US, CA 277812

[9] Lebedev Physical Institute, RAS, Moscow, Russia

[10] Institute of Acoustics, Chinese Academy of Sciences

[11] Institute of Atmospheric Physics, Chinese Academy of Sciences

* Corresponding author

**Abstract**

Due to the devastating consequences of major earthquakes, the identification of any reliable precursor signatures is of paramount importance. Some earthquakes indeed can be perceived several to tens of hours ahead. Although earthquake prediction is a big challenge in the world, some simple observational tools can capture many physical signals and demonstrate that an earthquake (EQ) may be forthcoming in short period. Many researchers have studied the significant variation of atmospheric electrostatic field related to the forthcoming earthquake[1-5]. However, until now, there is not a compelling physical mechanism which can explain why atmospheric electrostatic abnormal signal could appear just before an earthquake. Here we present a precursor signal and propose a brief physical interpretation. Under fair air conditions[6], if the near-surface atmospheric electrostatic field $E_z$ (oriented down when it is positive) presents a very stable negative anomaly (-100 V/m～-5000 V/m), it will forebode that an earthquake (seismicity from 3-8) would take place in the next several to tens of hours within a distance less than 100 km. We name this prediction technique as "DC $E_z$ Determination"(DED). In addition, the mechanism of such abnormal quasi-static electric field before a forthcoming earthquake has been proposed here: (1) Radon gas releases from the rock clefts near the epicenter during seismogenic process. (2) $α^+$ particles are produced due to the radioactive decay of Radon gas in the air (3) $α^+$ particle ionizing radiation creates more positive particles in the air, which is much more effective than that

β⁻ and γ particles produced during Radon radioactive decay. (4) The new positive particles change formal positive atmospheric electric field (Ez) into stably negative near the earth-surface. (5) The closer the instrument is to the epicenter, the more likely it is to observe the negative Ez signal related to the earthquake. It is recommended to establish an instrument network to capture the reliable precursor and make some warnings before the earthquake disaster.

In the search of a valid precursor for earthquake occurrence, we pose the question: Is there an hourly lasting atmospheric electrostatic field $E_z$ reversal before some earthquakes? The following examples (Fig. 1a, Fig. 1b and Fig. 2c) supply the evidence for a positive response to this question.

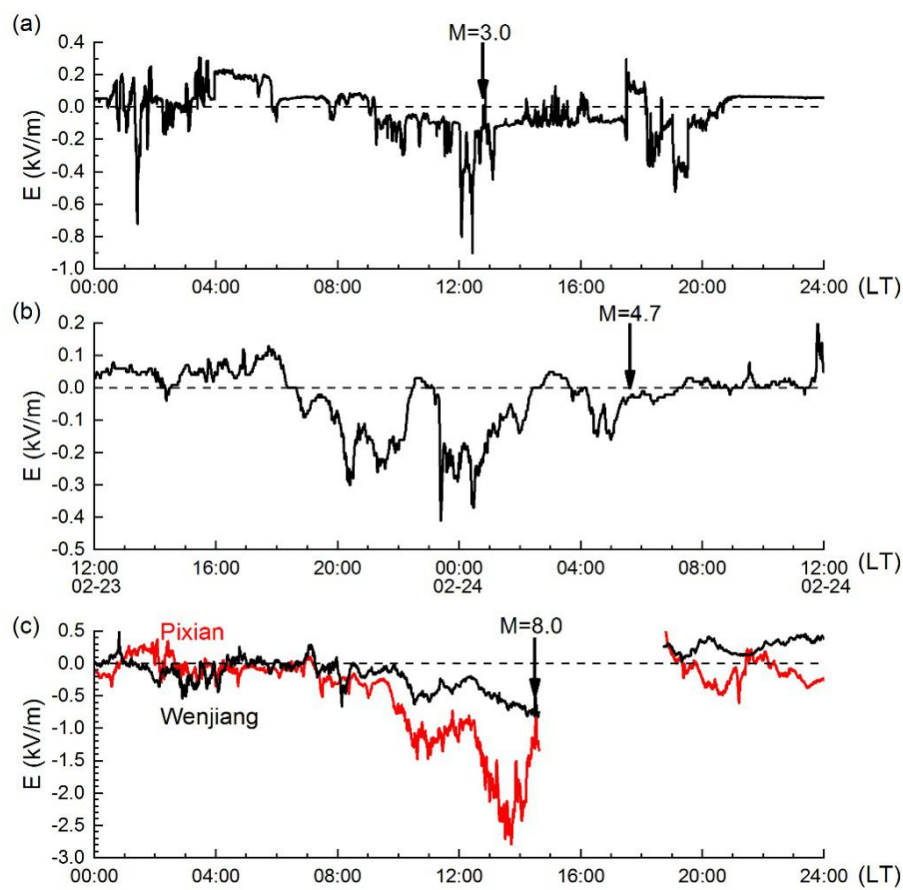

**Fig.1 Negative atmospheric electrostatic field $E_z$ before earthquakes.** (a) Beijing M3.0 earthquake on 14 April, 2019, with the station 40 km from the epicenter. (b) Rongxian M4.7 earthquake on 24 February, 2019, with the station 30 km from the epicenter. (c) Wenchuan M8.0 earthquake on 12 May, 2008, with two different stations 50 and 55 km from the epicenter.

The normal diurnal curve of atmospheric electrostatic field is called Carnegie curve[7], with a typical positive $E_z$ from tens of V/m to two hundreds of V/m. Due to the variability of solar activity, $E_z$ could be enhanced[8] to even 800 V/m.

On 14 April of 2019 in Beijing, China, the weather has been identified to be fair based on the criteria proposed by Harrison and Nicoll[6]. An abnormal phenomenon was observed when the normal positive $E_z$ (+100 V/m) changed into negative (about -100 V/m) around 9:00 LT in the morning, as shown in Fig. 1a. It is foreboded that a middle magnitude earthquake will be occur several hours later based on previous analytic results [4, 5, 9]. Later, a M3.0 earthquake was confirmed to occur at 12:47 LT with the epicenter 40 km north of the place that scientific instrument located.

Similarly, as shown in Fig. 1b, when the Rongxian M4.7 earthquake occurred at 05:38:10 on February 24, 2019 (Lon=104.49°, Lat=29.47°) in Sichuan Province, while at Zigong station situated 30 km from the epicenter, under fair air conditions the $E_z$ became negative about 12 hours ahead. The $E_z$ was suddenly reduced to a minimum of -410V/m at 23:24, in addition to two short intervals of positive $E_z$ (tens of minutes), the negative anomaly of electric field lasted for several hours until the earthquake began.

Another example is shown in Fig. 1c. In spite of the data gap, in Pixian county and Wenjiang county $E_z$ also acquired negative values for 7 hours before the Wenchuan M8.2 earthquake which occurred at 14:28 on May 12 of 2008. Note that the weather in Wenchuan county, Pixian county and Wenjiang county all the day is very fair as well. The peek negative value of $E_z$ at Pixian and Wenjiang station is -2750 V/m and -750 V/m, respectively. The Pixian station is 50 km off the epicenter, while the Wenjiang station is 55 km off the epicenter.

All the above examples indicate that there exists a negative atmospheric electrostatic field anomaly several hours before the occurrence of some earthquakes.

It is demonstrated that if the atmospheric electric field $E_z$ near earth surface shows a stable negative value (-100 V/m ~ -5000 V/m) under fair air conditions, meaning that without air pollution or sandy wind, an earthquake (M:3-8) will likely take place in the next several or tens hours within a distance less than 100 km.

The greater the intensity of EQ is, the more easily the reversal abnormality of the $E_z$ signal that would be applied for earthquake prediction can be recognized. The authors hope that more researchers from all over the world could pay attentions to the negative anomaly of $E_z$, especially under very fair weather conditions. Once seismogenic activity begin its latest period before earthquake, the released Radon gas from rock cleft heats the nearby atmosphere, reduces atmospheric humidity[10], and produces many positive charged particles in the air near the earth surface around the epicenter. Finally, a very strong and very stable upward electric field $E_z$ and current $J_z$ have been established. Because the $E_z$ signal strength is very obvious, it can be very useful to save many lives from the impending earthquake. Warning system can be easily devised to alert people to the abnormal signal and send alarm to retreat off dangerous place in emergency.

The above described perception method is named as "DC $E_z$ Determination" (DED).

Why is there an atmospheric electrostatic field reverse lasting for several hours just before some earthquakes? It is explained that the crust movement just before some earthquakes makes the rock cleft release significant amount of radioactive radon gas into the atmosphere near the Earth's surface, which further produces radiate $\alpha^+$ particles, $\beta^-$, $\gamma$ ray. Compared to $\beta^-$, $\gamma$ ray, $\alpha+$ particles have more powerful ionizing radiation capacity in the air. Then those $\alpha+$ particles will make more positive charged particles, whose electric field will exceed that of the negative charged particles previously existing near the ground. Finally, an electric field oriented upward purely would be established as shown in Fig. 2. The upward electric field not only compensates under fair weather conditions and the positive atmospheric electric field $E_z$ induced due to space weather, but also makes the observed $E_z$ present some negative signatures, which lasts for several or tens of hours in above cases. The mechanism of pre-earthquake related Radon gas inducing air ionization after its escape from the rocks near the epicenter region has been expressed by previous researchers [10-17]. The whole physical process of the hour-scale persistent negative anomaly of atmospheric electrostatic field near the epicenter before earthquake is shown in Fig. 2.

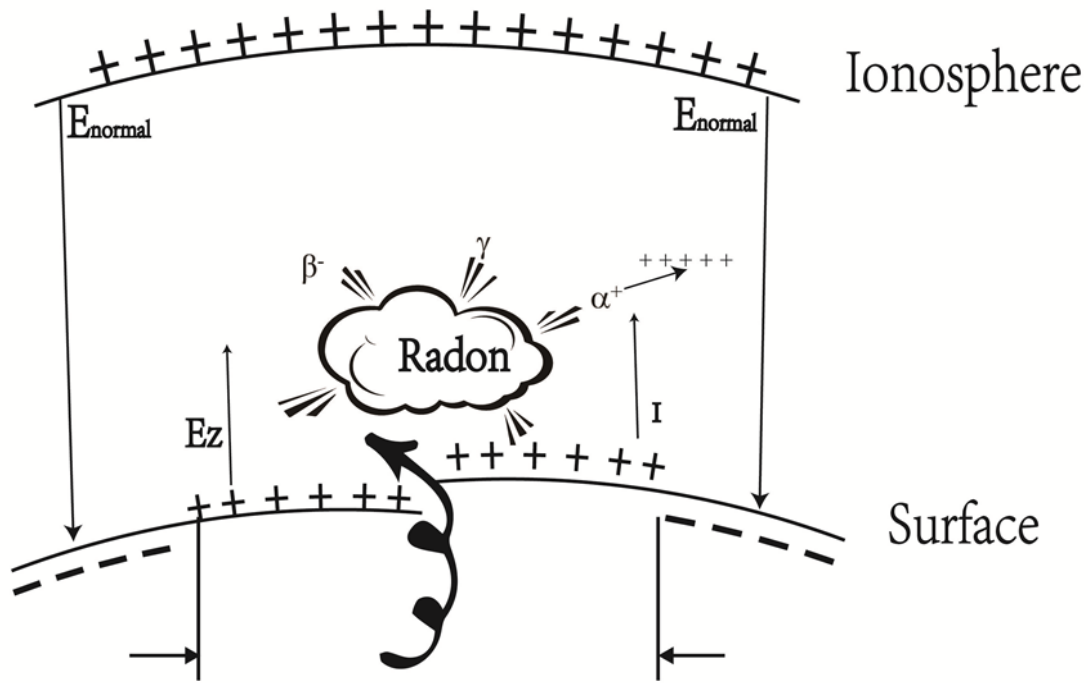

**Fig.2 Schematic diagram of the seismogenic process.** In fair weather, $E_{normal}$ presents the background electric field (The downward direction is defined as positive). $E_z$ and I (upward direction: negative) represent the electric field and current generated by a series of physical processes affected by the release of Radon gas. The later seismogenic process releases radon from rock cleft and produce $\alpha^+$ particle to complete ionizing radiation, so many positive charged particles to appear above the surface near the epicenter. Finally, a very strong electrostatic field has been established and lead to Ez negative signal stably. In general, the phenomenon lasts a few or tens hours just before earthquake.

Fig. 3 is an example that simulate a pre-earthquake electrostatic field appear in Wenchuan region and nearby just before the M8.0 earthquake. The electrostatic field may be inhomogeneous. The closer to the epicenter is, the greater the negative $E_z$ value measured by the instrument is. Therefore, before Wenchuan earthquake, the value of the $-E_z$ measured by Pixian station (50 km off the epicenter ) is greater (4 times) than that of Wenjiang station (55 km off the epicenter ), the two stations all showed that $-E_z$ appeared from 7'o clock till the earthquake occurred. The total $E_z$ negative anomaly lasted 7 hours (Fig. 1c). The comparison between the two simultaneous $E_z$ values from different observation sites demonstrates that the epicenter location must be nearby, perhaps less than 100 km of these observation sites.

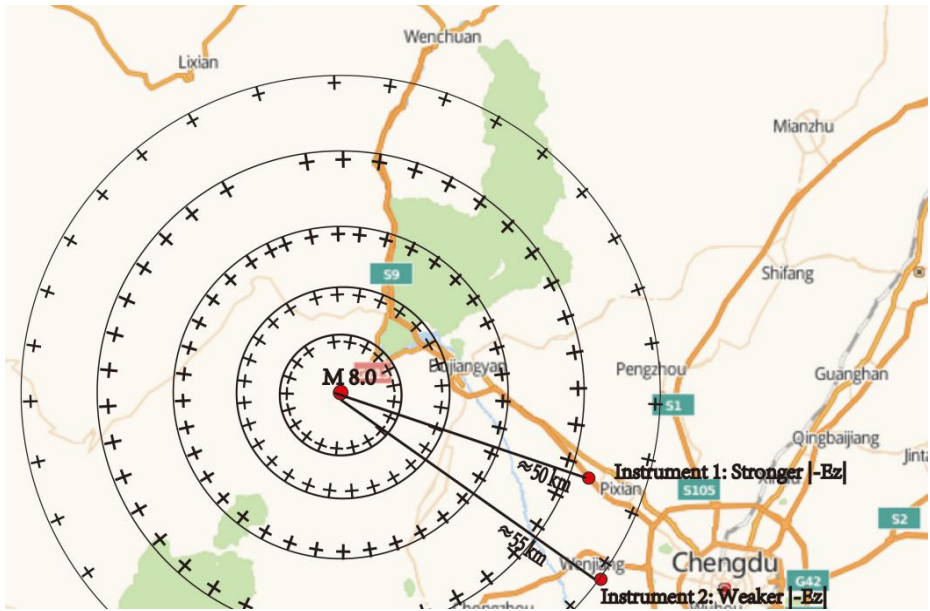

**Fig.3 A schematic diagram that simulates a pre-earthquake electrostatic field appear in Wenchuan region just before the M 8.0 earthquake.** The electrostatic field related earthquake has been observed by two stations in the meantime. The symbol + in the map illustrates the positive charge from Radon gas related to rock cleft due to seismogenic process.

It is suggested that once more stations would be established to monitor the $E_z$, more accurately one can determine what magnitude, when, and where the major earthquake. Geological condition, surface structure in epicenter region, the wind speed and direction at station during observation will affect the magnitude and lasting time of the $E_z$ negative value related to EQ. So the observed sites should be established according to localized pattern of the fault zones in the future.

The stronger an earthquake occurs, the greater the signal magnitude of the $E_z$ negative value may be. In general, the magnitude of a potential earthquake could be proportional to the magnitude of the negative magnitude of the $E_z$ according to localized geology condition.

It is concluded that:

1. Radon gas is released because of the dislocation of rocks just before earthquake. Radon gas is very radioactive, and it can ionize the atmosphere near the surface. Negative charges on surface are neutralized, so negative anomaly of atmospheric electric field is observed.

2. A very stable, lasting a few or tens hours atmospheric electrostatic vertical reverse electric $-E_z$ just before major earthquakes had been observed

3. The greater the earthquake (EQ) is, the greater the magnitude of the negative $E_z$ value is from the station near the epicenter due to a major EQ seismogenic process make more lastingly radon release to the air in more extensive region, then radon radioactive process make the $E_z$ value stable negative.

4. If in fair condition, the instrument perceives the $E_z$ value stable negative. The forthcoming earthquake will come nearby. In general, the distance is about less than 100 km.

5. The judgment method called "DC Ez Determination (DED)" has been presented

Tens of thousands people would die due to earthquakes every year (88 thousands people died during the M8.0 Wenchuan earthquake). If we can discern the seismology precursor even early over several to tens hours, it will benefit enormously to the humans near the earthquake epicenter. The above method can be applied by both non-experts and experts. The earthquake forecast researchers would like to use the atmospheric negative value (DC signal) abnormal signature to consult with meteorological experts and space weather experts to determine accurately the arrival time, class, influenced region range of some incoming earthquake. The authors sincerely hope that the "DC $E_z$ Determination (DED)" could help further promote the effectiveness of earthquake forecast. It will give enough time (several hours) to save millions of people with warning for leaving dangerous places before an earthquake disaster arrive.

**Methods**

The Fair weather conditions[6] which are used here for deleting any meteorological disturbance that may produce $E_z$ negative signal are as follows:

1) Lower relative humidity (No charged particles influenced by rain, snow or fog; less floating dust, snow; absence of aerosol and haze; visibility is greater than 2 km)

2) No obvious cumuliform and no plenty of stratus clouds with cloud base below 1.5 km.

3) The wind speed should less than 8 m/s within 10 meters of the surface.

Extended data Fig.1 displays the observations of atmospheric electric field under fair weather conditions. All the stations show a positive atmospheric electric field in a whole day under fair weather conditions. If a meteorological instrument is located near the electric field observation instrument, we can judge the cause of the negative anomaly to a certain extent. We test the negative anomaly observed in a week. Extended data Fig.2 show three cases of negative anomaly of atmospheric electric field. On the day of the earthquake (Shown in Extended Data Fig. 2a, and also in Text Fig. 1a), we observe that relatively high temperature, low relative humidity (before earthquake occurred at 12:47). And despite the higher wind speed, but lower PM10 index indicated that no significant dust that may affected the electric field during that day. Also, the stable positive signal between 7 o'clock and 9 o'clock before the earthquake indicates that the electric field value is not particularly disturbed by meteorological factors and human activities.

Compare with that, a very high relative humidity (Average value is greater than 80 %) can be observed on April 20, which indicated that the negative anomaly is due to the clouds or some rains. On April 21, the most characteristic is the electric field signal changed rapidly. During that day, high value of PM10 index was observed. And based on the statistical studies of electric field signal in Beijing by Ting et al.[18], the negative rapidly-changed value of atmospheric electric field can be explained as caused by sand particles and air pollutants.

**References (Of the methods)**

**Acknowledgements** This work was supported by the Strategic Pioneer Program on Space Science, Chinese Academy of Sciences, Grant No. XDA17010301, XDA17040505, XDA15052500, XDA15350201, and by the National Natural Science Foundation of China, Grant No. 41574161. The authors thank some supports from the Specialized Research Fund for State Key Laboratories, and CAS-NSSC-135 project. The authors also thank Prof. Fushan Luo, Prof. Jie Liu, Prof. Zhijun Niu and Prof. Shi Che for very useful discussion.

**Author contribution** Tao Chen designed the project. Han Wu, Xiaoxin Zhang, Xiaobing Jin, Lei Li and Jianguang Guo, Chong Wang collected the atmospheric electrostatic field $E_z$ data and analyzed them compared with the earthquake data. Professor Chi Wang, Jiyao Xu, Hui Li, Yong Liu, Suping Duan, Saiguan Xiao, Lei

Dai, Ilan Roth, Vladimir Makhumutov and Xiong Hu contribute space physics, atmospheric physics, atom physics theory and signal analysis. Qiming Ma, Jiajun Song, Fang Xiao, Jing Luo, Zhaohai He, Chen Zeng, and Xujie Jiang are responsible for building the atmospheric electric field instrument and calibration work and the $E_z$ measurement. Xichen Wang, Xuhui Shen, Quan Guo, Xiaodong Peng, Hanyin Cui, Chao Li, Qiang Sun provide sub-sonic and earthquake data and do some relative data analysis. Tao Chen wrote the paper with input from all authors.

**Competing interests** The authors declare no competing interests.

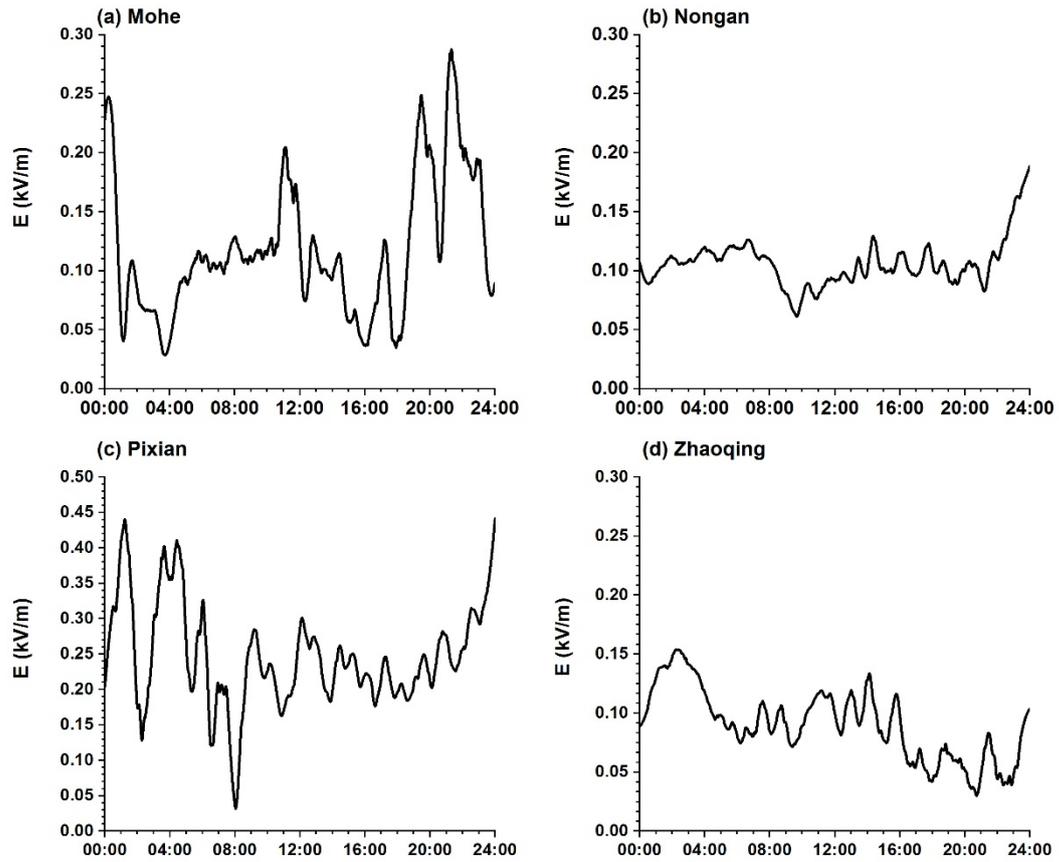

**Extended Data Fig.1** One-hour moving average of the atmospheric electric field in different stations of China under fair weather conditions.

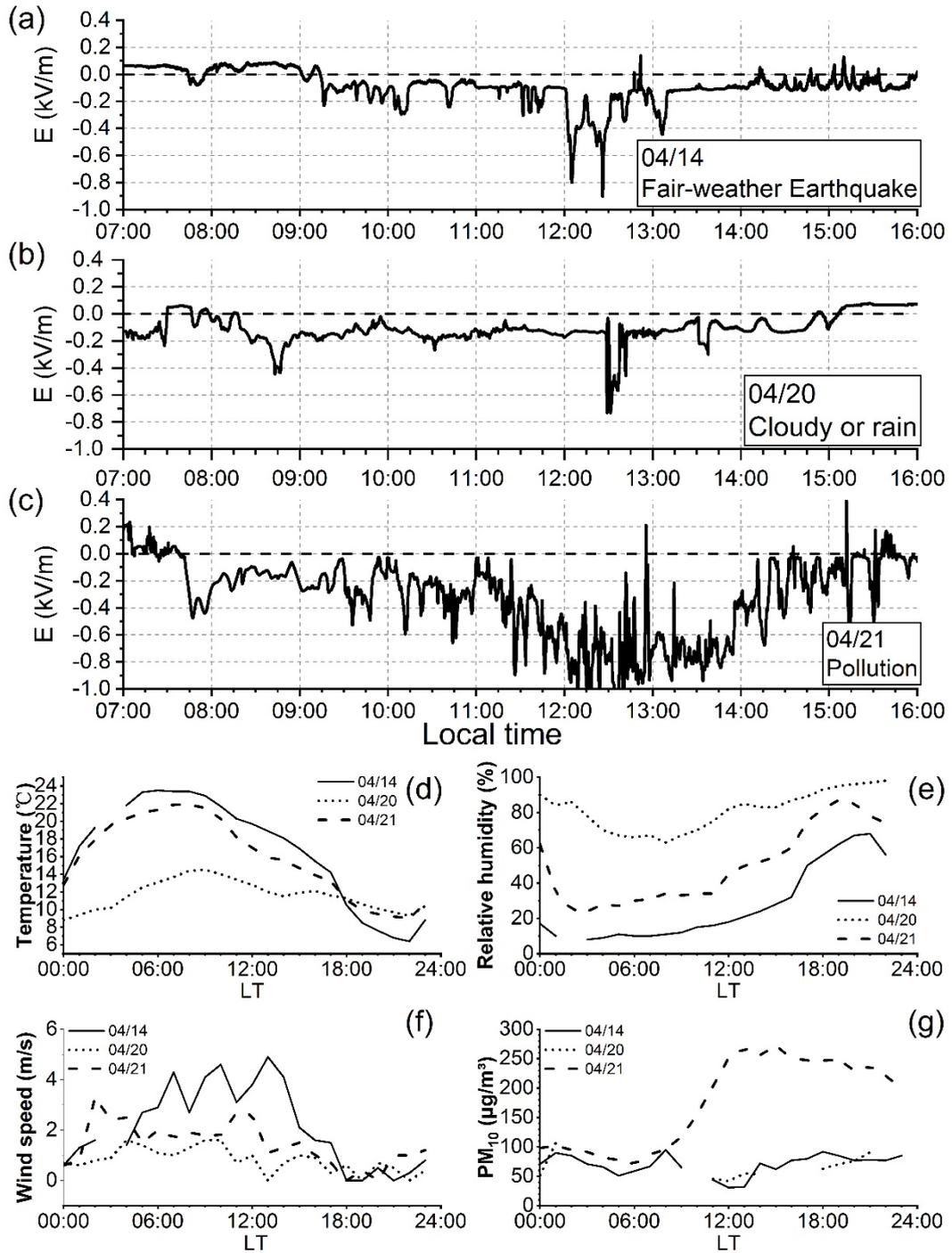

**Extended Data Fig.2 The comparison of coming seismology precursor characteristics with other weather negative Ez signal characteristics in 2019.** The meteorological observatories (provide temperature, wind speed, relative humidity elements) and pollution index monitoring instruments (provide PM 10 data) are within 5 km of the electric field instrument.